\pdfoutput=1
\documentclass[conference, 9pt]{IEEEtran}
\usepackage{amsmath,amssymb,amsfonts}
\usepackage{algorithmic}
\usepackage{graphicx}
\usepackage{textcomp}
\usepackage{xcolor}
\usepackage{siunitx}

\usepackage{tabularx}
\usepackage{pifont}
\usepackage{booktabs}

\usepackage{hyperref}
\usepackage[capitalize]{cleveref}
\usepackage{tikz}
\usepackage{pgfplots}
\pgfplotsset{compat=1.18}
\usepackage{bm}
\usepackage{balance}
\usepackage{upgreek}
\usepackage{multirow} % Allows rowspan
\usetikzlibrary{positioning, fit, calc}
\def\BibTeX{{\rm B\kern-.05em{\sc i\kern-.025em b}\kern-.08em
    T\kern-.1667em\lower.7ex\hbox{E}\kern-.125emX}}
\usepackage[acronym, shortcuts, toc,nonumberlist]{glossaries}

\newcommand{\gray}[1]{{\color{gray} #1}}

\newcommand{\mrow}[2][2]{\multirow{#1}[2]{*}{\begin{tabular}{@{}c@{}}#2\end{tabular}}}  % Centered row if using \cmidrule

\newcolumntype{H}{>{\setbox0=\hbox\bgroup}c<{\egroup}@{}}   % Hidden column https://tex.stackexchange.com/a/16607/148912
\glsdisablehyper    %% Disable hyperlinks for glossary entries. They don't make sense in papers.
\newacronym{cACGMM}{cACGMM}{complex Angular Central Gaussian Mixture Model}
\newacronym{GSS}{GSS}{Guided Source Separation}
\newacronym{ASR}{ASR}{Automatic Speech Recognition}
\newacronym{VMFMM}{vMFMM}{{von-Mises-Fisher Mixture Model}}
\newacronym{VMFcACGMM}{vMFcACGMM}{von-Mises-Fisher complex Angular Central Gaussian Mixture Model}
\newacronym{VAD}{VAD}{Voice Activity Detection}
\newacronym{DER}{DER}{Diarization Error Rate}
\newacronym{EER}{EER}{Equal Error Rate}
\newacronym{EM}{EM}{Expectation-Maximization}
\newacronym{WER}{WER}{Word Error Rate}
\newacronym{cpWER}{cpWER}{Concatenated minimum-Permutation Word Error Rate}
\newacronym{SLR}{SLR}{Segment Level Reassignment}

\usepackage{tikz}
\usetikzlibrary{arrows}
\usetikzlibrary{backgrounds}
\usetikzlibrary{fit}
\usetikzlibrary{positioning}
\usetikzlibrary{calc}
\usetikzlibrary{decorations.pathreplacing}
\usetikzlibrary{shapes.multipart}

\usetikzlibrary{shapes.misc}

\usepackage{tikzpagenodes}
\usetikzlibrary{tikzmark}
\usetikzlibrary{arrows.meta}  % https://tex.stackexchange.com/a/161238/148912

\usepackage{pgfplots}
\pgfplotsset{compat=1.13}

\tikzset{
    >=stealth,
    block/.style={rectangle,draw=black!100,fill=blue!25,minimum size=3em,text height=1.5ex,line width=0.1em,text depth=.25ex},
    branch/.style={circle,draw=black,fill=black,minimum size=0.25em,inner sep=0pt},
    arrow/.style={->, line width=0.1em},
    line/.style={-, line width=0.1em},
    reverse arrow/.style={<-,shorten <=0.1em},
}

\begin{document}

% \setlength{\abovecaptionskip}{1ex}
% \setlength{\belowcaptionskip}{1ex}
% \setlength{\floatsep}{1ex}
% \setlength{\textfloatsep}{1em}

% reduce the space above and below equations globally
\setlength{\abovedisplayskip}{4pt}
\setlength{\belowdisplayskip}{5pt}
% reduce the distance between floats on the top or the bottom and the text
\setlength{\textfloatsep}{5pt plus 0.0pt minus 2.0pt}
% reduce the distance between two floats
\setlength{\floatsep}{5pt plus 0.0pt minus 2.0pt}
% reduce the distance between two floats [h]
\setlength{\intextsep}{5pt plus 0.0pt minus 2.0pt}
% reduce the distance between columns in a table
% Tables according to Chicago Manual of Style

\title{Simultaneous Diarization and Separation of Meetings \\ through the Integration of Statistical Mixture Models}

\author{\IEEEauthorblockN{Tobias Cord-Landwehr, Christoph Boeddeker, Reinhold Haeb-Umbach}
\IEEEauthorblockA{\textit{Paderborn University, Communications Engineering Department, Paderborn, Germany}}}

\maketitle
\begin{abstract}
%%Tobias, suche Du aus, welche Version Dir besser gefaellt:
We propose an approach for simultaneous diarization and separation of meeting data.
It consists of a complex Angular Central Gaussian Mixture Model (cACGMM) for speech source separation, and a von-Mises-Fisher Mixture Model (vMFMM) for diarization in a joint statistical framework.
Through the integration, both spatial and spectral information are exploited for diarization and separation.
We also develop a method for counting the number of active speakers in a segment of a meeting to support block-wise processing.
While the total number of speakers in a meeting may be known, it is usually not known on a per-segment level.
With the proposed speaker counting, joint diarization and source separation can be done segment-by-segment, and the permutation problem across segments is solved, thus allowing for block-online processing in the future.
Experimental results on the LibriCSS meeting corpus show that the integrated approach outperforms a cascaded approach of diarization and speech enhancement in terms of WER, both on a per-segment and on a per-meeting level.

\end{abstract}

\section{Introduction}
\label{sec:intro}

Meeting transcription can be divided into three tasks: Diarization, which provides the annotation about who spoke when, and which consists of the subtasks segmentation and speaker identification; separation and enhancement which is about improving the quality and/or intelligibility of the speech signals;  and finally automatic speech recognition (ASR), the actual transformation of the speech into written text. 
Earlier works mainly focused on optimizing a single of these components to construct a modular pipeline, e.g.\ by solely focusing on the separation of overlapping speech or the diarization of meetings \cite{20_Chen_libricss, 21_rasj_meeting_processing}.
Recent works consider the complete pipeline as a holistic system \cite{23_yousefi_tsot_dia, 24_cornell_e2e_conv}, where all 
%processing 
steps are viewed as a single entity.

A wide-spread approach in meeting processing is \gls{GSS} \cite{18_boeddeker_gss}. 
First, diarization is performed on a meeting.
In essence, any system ranging from embedding-based systems using d-vectors \cite{17_li_deep_embeddings} or x-vectors \cite{18_snyder_xvector}, over neural speaker diarization systems \cite{23_plaquet_pyannote_powerset_eend} to multi-stage systems like TS-VAD \cite{20_medennikov_tsvad, 23_wan_ustc_chime7} can be used for diarization. 
This diarization is then used as a guide for an utterance-wise speech enhancement for the speech detected by the diarization.
To this end, a spatial mixture model that enhances the intelligibility and audio quality is employed for each utterance individually.
While \gls{GSS} allows for a robust speech enhancement and is employed in state-of-the-art systems \cite{23_cornell_chime7,23_wan_ustc_chime7,23_kamo_ntt_chime7,24_cornell_chime8} it builds upon previously detected segments and is unable to compensate for missed speech segments and speaker confusions.
Thus, diarization errors are propagated to subsequent processing stages due to the cascaded nature of the pipeline structure.
Other approaches are based on foregoing the guide and directly employing a \gls{cACGMM}, the underlying statistical mixture model of \gls{GSS}, for speech enhancement \cite{22_boeddeker_cacgmm_meetings}, or using a neural network-based approach also for the speech enhancement \cite{24_boeddeker_tssep,24_taherian_ssnd}. 
However, the statistical mixture-model-based approaches depend on accurate initialization and knowledge about the number of active sources.
Neural-network-based approaches like TS-SEP \cite{24_boeddeker_tssep} aim to refine an initial diarization estimate but require in-domain training data. 

In this work, we propose a model-based approach by integrating two models using spectral and spatial features for the diarization and separation components.
This is achieved by incorporating statistical mixture models designed for each task into a common statistical model that can be jointly fitted and optimized to the data. Here, a \gls{VMFMM} and a \gls{cACGMM} are integrated into a common statistical mixture model, the \gls{VMFcACGMM}.
The \gls{VMFMM} utilizes frame-level speaker embeddings to perform a diarization by grouping closely related speaker embeddings into the same cluster. 
The \gls{cACGMM} uses multi-channel STFT features to perform source separation and speech enhancement based on spatial cues by estimating the speech presence masks of each speaker.
The mixture models are coupled through their posterior probabilities so that the integrated model can exploit inter-dependencies between spectral and spatial cues in this way.
In addition, the speaker embeddings ease block-wise processing, as these speaker embeddings allow the re-identification of speakers between processed segments and can be used for speaker counting in each processed segment.

A related work is \cite{17_drude_integration} which also developed a joint spatial and spectral mixture model, however for the integration of two source separation models for short recordings. Compared to this, the proposed approach also encompasses diarization and source counting for the processing of meeting data.

This work is structured as follows.
First, \cref{sec:mm_meetings} briefly describes how statistical mixture models and the \gls{EM} algorithm are applied for meeting data processing.
In particular, a \gls{cACGMM}-based source separation and a \gls{VMFMM}-based diarization system are explained.
\Cref{sec:vmfcacgmm} then explains, how both statistical mixture models are combined into a common integrated approach.
Also, an approach for speaker counting utilizing the spectral diarization model is motivated.
In \cref{sec:eval}, the integrated model is evaluated against other meeting transcription models on the LibriCSS database in terms of segment- and meeting-level transcription performance.

% \newpage
\section{Mixture Model-based meeting processing}
\label{sec:mm_meetings}

% Vorschlag (im aktuellen text sind ein paar ungenauigkeiten):
We propose an integration of diarization and separation by modeling them with a joint spectral and spatial mixture model.
Then, the \gls{EM} algorithm is used for joint estimation of the parameters.

In general, the \gls{EM} algorithm aims to infer latent, unobservable variables $\mathcal{Z}$ from a set of observations $\mathcal{X}$.
For mixture models, the complete likelihood function of the observations is assumed to be the weighted product of $K$ distributions
\begin{align}
\label{eq:likelihood}
    \mathcal L(\mathcal{X}, \mathcal{Z}; \mathbf{\Theta}) = \prod_n \prod_k \left(\pi_k  p(\mathbf{x}_n; \Theta_k)\right)^{z_{k,n}},
\end{align}
where
% $k$ and $n$ are the mixture component index and 
the weight $\pi_k$ is the a prior probability $p(z_{k,n})$ of the hidden categorically distributed variable $z_{k,n} \in [0,1]$ denoting whether an observation  $n$ is drawn from mixture component $k$ \cite{McLachlan1996TheEA}.
The \gls{EM} algorithm works by alternately estimating the posterior probability ${\gamma_{k,n} = \mathrm{E}[z_{k,n}| \mathbf{x}_n ; \mathbf\Theta]}$
    % ${\gamma_n = E[z_n]}$
in the Expectation step (E-step) and then optimizing the model parameters $\mathbf\Theta$ to maximize the likelihood in the Maximization step (M-step).

For application in meeting processing, each active source is modeled by a mixture component. For instance, these components can represent speakers or positions in a room. Then, the mixture model is fit to the observations to estimate $z_{k,n}$.

\subsection{Spatial source separation model}
\label{sec:cacgmm}
\glsreset{cACGMM}
The \gls{cACGMM} \cite{16_ito_cacg_separation} is a statistical mixture model for source separation utilizing spatial cues. 
The input features are  $C$-dimensional microphone array observations $\mathbf{y}_{t,f}$ in the STFT domain, where $t$ and $f$ are the time frame and frequency bin index, respectively.
After normalizing the observation to unit length, these normalized observations $\tilde{\mathbf{y}}_{t,f}$ are assumed to be distributed as a mixture of complex Angular Central Gaussian distributions
\begin{align}
p(\tilde{\mathbf{y}}_{t,f}| z_{k,t,f}{=}1; \mathbf{B}_{k,f} )= \frac{(C-1)!}{2\pi^C\det{\mathbf{B}_{k,f}}}\frac{1}{\tilde{\mathbf{y}}_{t,f}^\mathrm{H}\mathbf{B}_{k,f}^{-1}\tilde{\mathbf{y}}_{t,f}},
\end{align}
which model the spatial cues through the frequency-dependent, spatial covariance matrices $\mathbf{B}_{k,f}$. 
% for a speech mixture of $K$ sources. 
% Here, the quadratic form \cref{eq:} is inferred from the last M-step since the update equation becomes intractable otherwise. 
After convergence, the result of the 
last 
E-step, the component posterior probabilities 
% $\mathrm{E}[z_{k,t,f}]=\gamma^{\mathrm{spat}}_{k,t,f}$ \fromcb{$= p(z_{k,t,f} | \tilde{\mathbf{y}}_{t,f} ; \mathbf\Theta^{\mathrm{spat}})$}
${\gamma^{\mathrm{spat}}_{k,t,f}= \mathrm{E}[z_{k,t,f} | \tilde{\mathbf{y}}_{t,f} ; \mathbf\Theta^{\mathrm{spat}}]}$
can be interpreted as time-frequency masks for each active source. 
Typically, these masks are used to perform mask-based beamforming \cite{15_yoshioka_nttchime3}. 

\subsection{Spectral diarization model}
\label{sec:vmfmm}
\glsreset{VMFMM}
For diarization, a single-channel \gls{VMFMM} using frame-level speaker embeddings as proposed in \cite{23_cordlandwehr_framewise_embeddings} is used.
First, a frame-level speaker embedding extractor is applied to the first microphone channel of the observation to obtain the speaker embeddings $\mathbf{e}_t$.
Whether two embeddings represent the same speaker is determined by the angle between the embeddings.
So, the length-normalized speaker embeddings $\tilde{\mathbf{e}}_t$ that lie on an $E$-dimensional hypersphere are used as observations for the spectral mixture model.
Therefore, the  von-Mises-Fisher distribution \cite{05_banerjee_vmf}
\begin{align}
p(\tilde{\mathbf{e}}_t | z_{k,t}{=}1; \varkappa_k, \boldsymbol{\mu}_k ) &= c_E(\varkappa_k)\exp\{{\varkappa_k\boldsymbol{\mu}_k^\mathrm{T}\tilde{\mathbf{e}}_t}\} \\
	c_E(\varkappa_k) &= \frac{\varkappa_k^{E/2 -1}}{(2\pi)^{E/2}I_{E/2}(\varkappa_k)} 
\end{align}
with the regular first-order Bessel function  $I_{E/2}(\cdot)$ of degree $E/2$ can be used to model the distribution of frame-level speaker embeddings for each speaker $k$ in a meeting.
Each mixture component is described by its orientation, which can be seen as an average speaker embedding, $\boldsymbol\mu_k$, and its concentration on the hypersphere $\varkappa_k$ in order to represent a single speaker. 
After convergence, the posterior probabilities $\gamma^{\mathrm{spec}}_{k,t}=  \mathrm{E}[z_{k,t}| \mathbf{e}_t, \boldsymbol \Theta_{\mathrm{spec}}]$  denoting the probability of speaker $k$ being active in a time frame $t$, can be used as a diarization estimate after smoothing and thresholding.

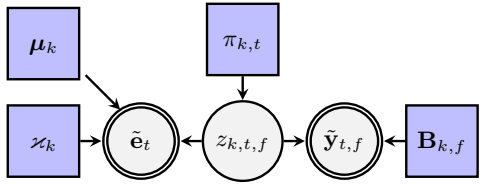
\begin{figure}[bt]
\centering
\vspace{-0.5em}
    \begin{tikzpicture}[
		node distance=1em,
		remember picture,
		% transform canvas={xshift=\linewidth - 1em, yshift=-1em}
	]
		\node (r) [block] {$\boldsymbol \mu_{k}$};
		\node (kappa) [block, below=of r] {$\varkappa_{k}$};
		\node (mix) [circle, minimum size=3em, text height=1.5ex, draw=black, line width=0.1em, fill=gray!10, double, right=of kappa] {$\tilde{\mathbf{e}}_t$};
		\node (mask) [circle, minimum size=3em, text height=1.5ex, line width=0.1em, draw=black,  fill=gray!10, right=of mix] {$z_{k,t,f}$};
		\node (pi) [block, line width=0.1em, above=of mask] {$\pi_{k,t}$};
		
		\draw[arrow] ($(r.south east) + (2pt, 2pt)$) --  ($(mix.north west) + (2pt, 2pt)$);
		\draw[arrow] (kappa) edge node {} ($(mix.west) + (-1pt, 0pt)$);
		\draw[arrow] (mask) edge node {} ($(mix.east) + (1pt, 0pt)$);
		\draw[arrow] (pi) edge node {} (mask);

		\node (obs) [circle, minimum size=3em, text height=1.5ex, draw=black, line width=0.1em, fill=gray!10, double, right=of mask] {$\tilde{\mathbf{y}}_{t,f }$};
  		\node (sigma) [block, right=of obs] {$\mathbf{B}_{k,f}$};
            \draw[arrow] (mask) edge node {} ($(obs.west) + (-1pt, 0pt)$);
		\draw[arrow] (sigma) edge node {} ($(obs.east) + (1pt, 0pt)$);
\end{tikzpicture}
    \vspace{-0.5em}
    \caption{Graphical model of the integrated mixture model. The spectral diarization  model (left) and the spatial speech enhancement model (right) are coupled through the common latent state $z$.}
    \label{fig:vmfcacgmm}
\end{figure}

\section{Integration of separation and diarization}
\label{sec:vmfcacgmm}
While both models can be individually applied to meeting data, they both suffer from drawbacks. The \gls{cACGMM} only handles spatial cues, so nearby speakers can be confused and the model can split a single speaker into multiple mixture components due to strong reflections in the room \cite{22_boeddeker_cacgmm_meetings}. 
On the other hand, the \gls{VMFMM} has been shown to mainly achieve a good diarization performance for single-speaker regions. For overlapping speech, the mixture model needs to be carefully tuned to represent multiple active speakers, at all. 
This is crucial since this approach actually violates the sparsity assumption in \cref{eq:likelihood}: The latent variable can no longer be assumed to be categorically distributed if a frame-level embedding $\mathbf{e}_t$ is to represent the mixture of multiple speakers. 
To mitigate these weaknesses, the spatial \gls{cACGMM} and spectral \gls{VMFMM} are integrated into a common mixture model by coupling them 
through the latent variable $z_{k,t,f}$.
As in \cite{17_drude_integration}, this is done by assuming conditional statistical independence 
\begin{align}
\begin{split}
p(\tilde{\mathbf{y}}_{t,f}, \tilde{\mathbf{e}}_t | z_{k,t,f}{=}1; \mathbf{B}_{k,f}, \varkappa_k, \boldsymbol{\mu}_k) =  \\ p(\tilde{\mathbf{y}}_{t,f} | z_{k,t,f}{=}1; \mathbf{B}_{k,f})p(\tilde{\mathbf{e}}_t |z_{k,t,f}{=}1; \varkappa_k, \boldsymbol{\mu}_k)
\end{split}
\end{align}
between spectral and spatial variables.
To do so, the \mbox{von-Mises-Fisher} distribution is replicated along the frequency axis, as this axis is newly introduced for the spectral model. The graphical model of the integrated mixture model is given in \cref{fig:vmfcacgmm}.
Using the independence assumption results in the log-likelihood   
\begin{align}
\begin{split}
\ell(\tilde{\mathbf{y}}_{t,f}, \tilde{\mathbf{e}}_t; \mathbf{\Theta}) &= \sum_t \sum_f\sum_k z_{k,t,f}(\ln(\pi_{k,t}) \\\phantom{=}+ &\ln(p(\tilde{\mathbf{y}}_{t,f}; \mathbf{B}_{k,f})) + \ln(p(\tilde{\mathbf{e}}_{t}; \varkappa_{k}, \boldsymbol{\mu}_k)))
\end{split}
\end{align}
for the integrated model according to \cref{eq:likelihood}.
As proposed in \cite{13_ito_frequency_tied,22_boeddeker_cacgmm_meetings}, the a-priori probabilities $\pi_{k,t}$ of each component are set frequency-independent, but time-dependent which allows to model a time-varying speech presence probability of each speaker.
%in a meeting.
In the E-step, the latent variable $z_{k,t,f} $ is replaced with its expected value 
\begin{align}
    \gamma_{k,t,f} {=} \frac{\pi_{k,t}p(\tilde{\mathbf{y}}_{t,f}|z_{k,t,f}{=}1; \mathbf{B}_{k,f})p(\tilde{\mathbf{e}}_t|z_{k,t,f}{=}1; \varkappa_k,\boldsymbol{\mu}_{k})}{\sum\limits_{\tilde{k}}\pi_{\tilde{k},t}p(\tilde{\mathbf{y}}_{t,f}|z_{\tilde{k},t,f}{=}1; \mathbf{B}_{\tilde{k},f})p(\tilde{\mathbf{e}}_t|z_{\tilde{k},t,f}{=}1; \varkappa_{\tilde{k}},\boldsymbol{\mu}_{\tilde{k}})}.
\end{align}
After replacing $z_{k,t,f}$ with its estimate $\gamma_{k,t,f}$ in the log-likelihood (resulting in the auxiliary function $Q(\cdot)$), the 
parameters of the mixture components are updated to maximize the log-likelihood 
during the M-step. 
This results in an unchanged update formulation compared to the individual models because the gradients 
\begin{align}
    \nabla_{\boldsymbol\Theta_{\mathrm{spat}}}Q &= \nabla_{\boldsymbol\Theta_{\mathrm{spat}}} \sum_t\sum_f\sum_k \gamma_{k,t,f}\ln(p(\tilde{\mathbf{y}}_{tf};\mathbf{B}_{k,f})) \\
    \nabla_{\boldsymbol\Theta_{\mathrm{spec}}}Q  &= \nabla_{\boldsymbol\Theta_{\mathrm{spec}}} \sum_t\sum_k\bar{\gamma}_{k,t}\ln(p(\tilde{\mathbf{e}}_{t};\varkappa_k,\boldsymbol{\mu}_{k}))
\end{align}
w.r.t.\ the mixture model parameters for the spatial model $\boldsymbol\Theta_{\mathrm{spat}}$ and spectral model $\boldsymbol\Theta_{\mathrm{spec}}$ are independent of each other and the update equations from \cite{05_banerjee_vmf} and \cite{87_tyler_cacg} can be used.
Here, $\bar{\gamma}_{k,t}$ is the sum of the posteriors $\gamma_{k,t,f}$ over all frequencies, which is used as the expected value of the latent variable during the M-step of the \gls{VMFMM}.

From these derivations, it can be seen that integrating the \gls{VMFMM} with the \gls{cACGMM} into the \gls{VMFcACGMM} aims to compensate for the weaknesses of the individual models. 
First, spectral cues are introduced without complicating the individual update equations of either mixture model compared to a \gls{cACGMM}-based speech enhancement. 
Second, the availability of a frequency dimension in the spectral model eases the detection of more than one concurrent speaker, improving its capability to handle overlapping speech.
This is achieved since by averaging over the frequency components, the class posteriors used for the \gls{VMFMM} parameter update depict a linear combination of all active classes in a time frame.

\subsection{Embedding-based component fusion \& speaker counting}
\label{sec:fusion}
When dividing a meeting into segments to reduce memory and processing latency, or to account for moving sources, the number of active speakers in a segment is usually unknown, even if the total number of participants in a meeting is known.
Not only needs the number of speakers in a segment be estimated, also the assignment of speakers between the segments, the segment permutation problem, must be solved to re-identify speakers in different segments.

Integrating the \gls{cACGMM} with the \gls{VMFMM} allows for an easy component fusion.
Each mixture component $k$ of the integrated \gls{VMFcACGMM} is specified by its spatial cues represented in $\mathbf{B}_{k,f}$, and the spectral cues in terms of its spread of speaker embeddings $\varkappa_k$ and the average orientation $\boldsymbol\mu_k$. This average orientation can also be interpreted as a prototype speaker embedding of the mixture component.
Therefore, the component fusion can be designed similarly to the objective of speaker verification \cite{22_wang_cosine_sv}. 
By comparing the cosine similarity of all mixture components' average orientation, superfluous components can be detected, and split speakers can be merged.
First, all pairwise cosine similarities $s_{ij}$
between all mixture components $i,j$ are computed. 
If any similarities exceed the fusion threshold $s_{ij} > \tau$, the two mixture components with the highest similarity are fused before continuing the \gls{EM} algorithm.
During fusion, the posterior probabilities $\gamma$ of the two components are combined. The \gls{cACGMM} additionally requires the spatial covariance matrix of the previous step for optimization \cite{87_tyler_cacg},  so the new value 
\begin{align}
\gamma_{i,t,f} &= \gamma_{i,t,f} + \gamma_{j,t,f} \\
\mathbf{B}_{i,f} &=  \mathbf{B}_{i,f}\frac{\sum_t\pi_{i,t}}{\sum_t\pi_{i,t} + \sum_t\pi_{j,t}} + \mathbf{B}_{j,f}\frac{\sum_t\pi_{j,t}}{\sum_t\pi_{i,t} + \sum_t\pi_{j,t}}
\end{align}
is calculated as the weighted average of the fused mixture components and the mixture component $j$ is deleted afterwards.

\section{Experiments}
\label{sec:eval}

\subsection{Evaluation Setup}
All models are evaluated on the LibriCSS database \cite{20_Chen_libricss}, which consists of re-recordings of simulated meetings using LibriSpeech data with \SIrange{0}{40}{\percent} overlap. Here, results of both a segment-wise and a meeting-wise processing are reported, where segments are determined by speech pauses as specified in \cite{20_Chen_libricss}, ranging from \SIrange{30}{60}{\second} on average, while the full meeting is \SI{10}{\min} long. Results are given in terms of \gls{cpWER} using the MeetEval toolkit \cite{23_neumann_meeteval}. 

The observations $\mathbf{y}_{t,f}$ of the spatial mixture model consist of all \num{7} microphone channels, transformed into the STFT domain with an STFT size of \SI{64}{\milli\second}, a window size of \SI{50}{\milli\second}, and a frame advance of \SI{16}{\milli\second}, respectively.
For the frame-level embeddings $\mathbf{e}_t$ of the spectral model, the embedding extractor from \cite{24_cordlandwehr_geodesic} is used, which is trained on partially overlapping speech mixtures and outputs \num{64}-dimensional speaker embeddings with a frame-level resolution. 
This embedding extractor is trained in a teacher-student fashion, such that the frame-level embeddings aim to replicate ordinary speaker embeddings.
%A geodesic loss regularizes embeddings of overlapping speech to represent the combination of all active speakers.
The embedding extractor is trained using VoxCeleb mixtures, so no in-domain data from LibriSpeech or LibriCSS is used during the training phase.
For embedding-based component fusion (\textit{spectral}), a threshold of $\tau=0.7$ is chosen, for the intersection-over-union-ratio-based fusion (\textit{IoU}) \cite{22_boeddeker_cacgmm_meetings}, a threshold of $\tau=0.85$ is chosen.

All models use WPE \cite{18_drude_wpe} for preprocessing and a wMPDR beamformer \cite{19_nakatani_wmpdr} for speech extraction to match the experimental setup in \cite{22_boeddeker_cacgmm_meetings}. To ensure convergence, the mixture models are fitted for \num{100} EM steps.
Afterward, the priors $\pi_{k,t}$ are smoothed and used to detect individual utterances of each speaker, serving as the diarization estimate. The detected utterances are then beamformed and transcribed.
For \gls{ASR}, the ESPnet model from \cite{watanabe2020PretrainedASR} and the Nemo \gls{ASR} model \cite{23_rekesh_nemo_asr} are used.

\subsection{Mixture model initialization}
The \gls{VMFMM} from  \cite{24_cordlandwehr_geodesic} which serves as the spectral part of the \gls{VMFcACGMM} is used to initialize the mixture model. 
First, an energy-based \gls{VAD} using minimum statistics is applied to the recording. 
Then, the initialization, i.e.\ the diarization model, is run only on the voiced segments. 
A k-Means clustering \cite{07_arthur_kmeans} is applied to the frame-level embeddings $\mathbf{e}_t$ to obtain initial clusters. 
Then, these clusters are used as initialization of a \gls{VMFMM}, which is fitted for \num{30} iterations to the data. 
Here, the maximal concentration of each mixture component is limited to $\varkappa_{\mathrm{max}}=35$, which was shown to offer a good trade-off between accuracy and capability for overlap detection in \cite{24_cordlandwehr_geodesic}. 
Afterward, all silence regions detected by the \gls{VAD} are set active in an additional noise component. 
The posterior probabilities of this mixture model represent the soft-decision diarization estimates which are used as initial values for the integrated \gls{VMFcACGMM}.
To this end, the initialization values are replicated across the frequency axis and set as initial values for $\gamma_{k,t,f}$. The \gls{EM}-algorithm is started with an M-step to obtain the initial model parameters. 

For evaluation, a global initialization (\textit{VMF global}) by using the diarization from \cite{24_cordlandwehr_geodesic} to initialize each segment, and an initialization per segment (\textit{VMF}) with $K$ components per segment is compared.

\begin{table*}[bht]
    \sisetup{detect-weight}
	\robustify\bfseries  % https://tex.stackexchange.com/a/573662/148912
    \sisetup{round-precision=1,round-mode=places, table-format = 2.1}
% 	\sisetup{round-precision=2,round-mode=places, table-format = 1.2}\toprule
\vspace{-0.5em}
\caption{\gls{cpWER} of the \gls{VMFcACGMM} on LibriCSS segments compared to only using a \gls{cACGMM} as spatial mixture model. Models marked with * assume knowledge of the number of active speakers per segment.}
\label{tab:segment_results}
\centering
\vspace{-0.5em}
\begin{tabular}{c l l c S S S S S S S | S S}
\toprule
\mrow{row} &\mrow{Model} & \mrow{Initialization} &  \mrow{\shortstack[c]{Component \\ fusion}} & {\mrow{\shortstack[c]{DER}}} &  \multicolumn{7}{c}{{ESPnet ASR}} & {Nemo ASR}\\
\cmidrule(lr){6-12}\cmidrule(lr){13-13}
& & && &{0S} & {0L} & {OV10} & {OV20} & {OV30} & {OV40} & {avg.} & {avg.}\\
\midrule
    1 & VMF global \cite{24_cordlandwehr_geodesic} & k-Means++ &  -- & 10.4 & {--} &  {--} &{--}  & {--} & {--} & {--} & {--} & {--}\\
    2 &cACGMM & VMF global \cite{24_cordlandwehr_geodesic} & IoU & 18.7 & 9.942 & 14.3957 & 13.745 & 16.409 & 17.982 & 19.267 & 15.6 & 14.7\\% /scratch/hpc-prf-nt1/cord/models/vmfcacgmm_segments/spotty_green_parrot
    3 & cACGMM* \cite{22_boeddeker_cacgmm_meetings} & cACG  \cite{22_boeddeker_cacgmm_meetings} & IoU & {--} & 4.2 & 4.3 &  4.2 & 5.7 & 7.3 & 8.5 & 5.9 & {--}\\ 
        % 3 & cACGMM & VMF dia \cite{24_cordlandwehr_geodesic} & spectral & & & & & & & 7.4 \\ % /scratch/hpc-prf-nt1/cord/models/vmfcacgmm_segments/spotty_green_parrot_spectral_fusion
    4 & \gray{cACGMM*} & \gray{oracle} & \gray{--} & \gray{\hphantom{1}6.4}&  \gray{\hphantom{1}3.9} & \gray{\hphantom{1}4.0} & \gray{\hphantom{1}4.2} & \gray{\hphantom{1}5.3} & \gray{\hphantom{1}6.5} & \gray{\hphantom{1}6.1} & \gray{\hphantom{1}5.1} & \gray{\hphantom{1}4.7} \\  %/scratch/hpc-prf-nt1/cord/models/vmfcacgmm_segments/vivacious_bronze_roundworm
    \midrule
    5 & cACGMM & VMF global \cite{24_cordlandwehr_geodesic} & spectral & 13.1 & 13.7 & 9.3 & 6.3& 7.8 & 10.0 & 12.5 & 8.7 & 7.4\\
    6 & VMFcACGMM & VMF global \cite{24_cordlandwehr_geodesic}  & spectral &10.8 & 6.479 & 11.312 & 5.867 &6.545  & 8.322 & 10.828 & 8.2& 6.8\\ %/scratch/hpc-prf-nt1/cord/models/vmfcacgmm_segments/high_brown_bonobo
    % 6 &VMFcACGMM & random (K=8) & spectral & & & & & & & 9.49126025534768\\
 %/scratch/hpc-prf-nt1/cord/models/vmfcacgmm_segments/live_coffee_primate
    % VMFcACGMM & VMF (K=12) & spectral & & & &  & & & 6.694728215825596\\
    7 & VMFcACGMM & VMF (K=8) & spectral & 9.6& 6.3483 & 7.5133 & 4.426 & 6.189 & 8.925 & 10.055 & 7.3596 & 6.1\\ %/scratch/hpc-prf-nt1/cord/models/vmfcacgmm_segments/conservation_maroon_swan   
    8 & VMFcACGMM & VMF (K=10) & spectral &9.7  &5.5913 & 7.546 & 4.2918 & 5.83723& 7.6275& 9.6849& 6.8418  & 5.7  \\ 
    
    9 & VMFcACGMM* & VMF (K=8)   & spectral  & 8.2 & 4.523  & 6.232 & 4.176  & 5.428 &7.112 & 7.615&  5.8 & 5.4\\ %/scratch/hpc-prf-nt1/cord/models/vmfcacgmm_segments/crooked_azure_stoat
    10 & \gray{VMFcACGMM*} & \gray{oracle} & \gray{--} & \gray{\hphantom{1}6.4}& \gray{\hphantom{1}4.4} &\gray{\hphantom{1}5.7}  & \gray{\hphantom{1}3.6} & \gray{\hphantom{1}4.3} & \gray{\hphantom{1}5.6} & \gray{\hphantom{1}5.3} & \gray{\hphantom{1}4.9}&\gray{\hphantom{1}4.3}\\ %/scratch/hpc-prf-nt1/cord/models/vmfcacgmm_segments/vivacious_bronze_roundworm
    \bottomrule
\end{tabular}
\vspace{-2ex}
\end{table*}

\subsection{Segment-level Results}
\label{sec:segment_result}
First, the integrated \gls{VMFcACGMM} is compared to only using a spectral or spatial mixture model, on the LibriCSS segments. This evaluation only considers the \gls{WER} on each LibriCSS segment, containing \SIrange{30}{60}{\second}, without taking into account any segment permutation problems. 
\Cref{tab:segment_results} shows that only using the \gls{VMFMM} diarization (row \num{1}) to initialize a \gls{cACGMM} (row \num{2}) leads to high \glspl{WER}.
This is mainly due to diarization errors leading to missed speakers and multiple mixture components collapsing to the same speaker.
A component fusion based on the intersection of component priors (IoU) as done in \cite{22_boeddeker_cacgmm_meetings} proves unable to compensate for these errors.
If the number of speakers in a segment is known, a careful initialization of the \gls{cACGMM} combined with the IoU-based fusion can be designed to obtain a good \gls{WER} (row \num{3}). Compared to row \num{2}, using the frame-level speaker embeddings for a spectral component fusion already helps to significantly improve the performance compared to the prior-based fusion (row \num{5}). The \gls{VMFcACGMM} then additionally improves upon this 
(row \num{6}).  
This shows that the additional availability of spectral information is beneficial for the mixture model.
By initializing the mixture model segment-by-segment 
instead of using the diarization estimate of the whole meeting, the performance can be further improved (rows \numrange{7}{9}).
Here, starting the EM iterations with a number of components $K$ equal to or above the total number of speakers in the meeting and using the spectral component fusion of \cref{sec:fusion}, the \gls{WER} reduces to \SI{6.8}{\percent} (ESPnet, rows \num{7}/\num{8}).
If the total number of classes is used as a stopping criterion for the fusion (row \num{9}), a \gls{WER} of \SI{5.8}{\percent} is achieved, outperforming the \gls{cACGMM} from \cite{22_boeddeker_cacgmm_meetings} and reducing the gap to using an oracle diarization as initialization (row \num{10}).

\subsection{Speaker counting performance}
\begin{figure}
    \centering
        \vspace{-0.75em}
    % This file was created with tikzplotlib v0.10.1.
\begin{tikzpicture}

\definecolor{darkgray176}{RGB}{176,176,176}

\begin{axis}[
% axis background/.style={fill=lavender234234242},
% axis line style={white},
% colorbar,
% colorbar style={ylabel={}},
colormap/blackwhite,
point meta max=75,
point meta min=0,
tick align=outside,
xtick pos=left,
ytick pos=left,
x grid style={white},
xlabel=\textcolor{black}{Estimated speakers},
xmajorgrids,
xmajorticks=true,
xmin=-0.5, xmax=7.5,
xtick style={color=black},
xtick={0,1,2,3,4,5,6,7},
xticklabels={1,2,3,4,5,6,7,8},
y dir=reverse,
y grid style={white},
ylabel=\textcolor{black}{Active speakers},
ymajorgrids,
ymajorticks=true,
ymin=-0.5, ymax=7.5,
ytick style={color=black},
ytick={0,1,2,3,4,5,6,7},
yticklabels={1,2,3,4,5,6,7,8},
height=5cm
]

\tikzstyle{scaNumber}=[scale=0.7, text=black, rotate=0.0]
\tikzstyle{scaNumberWhite}=[scale=0.7, text=white, rotate=0.0]

\addplot graphics [includegraphics cmd=\pgfimage,xmin=-0.5, xmax=7.5, ymin=7.5, ymax=-0.5] {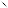};
\draw (axis cs:0,0) node[scaNumber]{8};
\draw (axis cs:1,0) node[scaNumber]{0};
\draw (axis cs:2,0) node[scaNumber]{0};
\draw (axis cs:3,0) node[scaNumber]{0};
\draw (axis cs:4,0) node[scaNumber]{0};
\draw (axis cs:5,0) node[scaNumber]{0};
\draw (axis cs:6,0) node[scaNumber]{0};
\draw (axis cs:7,0) node[scaNumber]{0};
\draw (axis cs:0,1) node[scaNumber]{0};
\draw (axis cs:1,1) node[scaNumber]{34};
\draw (axis cs:2,1) node[scaNumber]{0};
\draw (axis cs:3,1) node[scaNumber]{0};
\draw (axis cs:4,1) node[scaNumber]{0};
\draw (axis cs:5,1) node[scaNumber]{0};
\draw (axis cs:6,1) node[scaNumber]{0};
\draw (axis cs:7,1) node[scaNumber]{0};
\draw (axis cs:0,2) node[scaNumber]{1};
\draw (axis cs:1,2) node[scaNumber]{2};
\draw (axis cs:2,2) node[scaNumberWhite]{77};
\draw (axis cs:3,2) node[scaNumber]{9};
\draw (axis cs:4,2) node[scaNumber]{0};
\draw (axis cs:5,2) node[scaNumber]{0};
\draw (axis cs:6,2) node[scaNumber]{0};
\draw (axis cs:7,2) node[scaNumber]{0};
\draw (axis cs:0,3) node[scaNumber]{0};
\draw (axis cs:1,3) node[scaNumber]{1};
\draw (axis cs:2,3) node[scaNumber]{8};
\draw (axis cs:3,3) node[scaNumberWhite]{151};
\draw (axis cs:4,3) node[scaNumber]{17};
\draw (axis cs:5,3) node[scaNumber]{0};
\draw (axis cs:6,3) node[scaNumber]{0};
\draw (axis cs:7,3) node[scaNumber]{0};
\draw (axis cs:0,4) node[scaNumber]{0};
\draw (axis cs:1,4) node[scaNumber]{0};
\draw (axis cs:2,4) node[scaNumber]{2};
\draw (axis cs:3,4) node[scaNumber]{12};
\draw (axis cs:4,4) node[scaNumberWhite]{153};
\draw (axis cs:5,4) node[scaNumber]{14};
\draw (axis cs:6,4) node[scaNumber]{0};
\draw (axis cs:7,4) node[scaNumber]{0};
\draw (axis cs:0,5) node[scaNumber]{0};
\draw (axis cs:1,5) node[scaNumber]{0};
\draw (axis cs:2,5) node[scaNumber]{0};
\draw (axis cs:3,5) node[scaNumber]{2};
\draw (axis cs:4,5) node[scaNumber]{17};
\draw (axis cs:5,5) node[scaNumberWhite]{109};
\draw (axis cs:6,5) node[scaNumber]{10};
\draw (axis cs:7,5) node[scaNumber]{0};
\draw (axis cs:0,6) node[scaNumber]{0};
\draw (axis cs:1,6) node[scaNumber]{0};
\draw (axis cs:2,6) node[scaNumber]{0};
\draw (axis cs:3,6) node[scaNumber]{0};
\draw (axis cs:4,6) node[scaNumber]{1};
\draw (axis cs:5,6) node[scaNumber]{13};
\draw (axis cs:6,6) node[scaNumber]{52};
\draw (axis cs:7,6) node[scaNumber]{1};
\draw (axis cs:0,7) node[scaNumber]{0};
\draw (axis cs:1,7) node[scaNumber]{0};
\draw (axis cs:2,7) node[scaNumber]{0};
\draw (axis cs:3,7) node[scaNumber]{0};
\draw (axis cs:4,7) node[scaNumber]{0};
\draw (axis cs:5,7) node[scaNumber]{0};
\draw (axis cs:6,7) node[scaNumber]{9};
\draw (axis cs:7,7) node[scaNumber]{12};
\end{axis}

\end{tikzpicture}
        \vspace{-1em}
    \caption{Speaker counting accuracy of the \gls{VMFcACGMM} in \cref{tab:segment_results}, row \num{7} 
    % on LibriCSS segments
    }
    \label{fig:speaker_counting}
\end{figure}
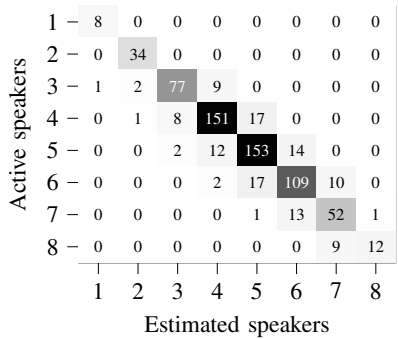

The \gls{VMFcACGMM} tries to automatically infer the number of active speakers in a segment according to the embedding-based component fusion specified in \cref{sec:fusion}.
\Cref{fig:speaker_counting} shows that while the mixture model can determine the active speakers correctly with an accuracy of \SI{84}{\percent}, i.e.\ \num{596} out of \num{725} segments, the errors mainly occur in segments with more than \num{3} active speakers.
Here, mainly low-activity speakers that completely intersect with other speakers are wrongly detected or merged, so that the \SI{16}{\percent} speaker errors only contribute \SI{1}{\percent} to the total \gls{WER} (compare ESPnet rows \num{8}/\num{9}).  Therefore, the embedding-based component fusion poses an efficient and simple speaker counting.

\subsection{Comparison of segment-level  and meeting-level results}
While the total number of speakers in a meeting can be assumed to be known, the number of active speakers within a short segment of the meeting will hardly be known.
Contrary to the \gls{cACGMM}, the \gls{VMFcACGMM} directly allows for segment-wise processing of meetings, because it can robustly estimate the number of speakers.
Further, the speaker prototypes $\boldsymbol\mu_k$ allow solving the segment permutation problem:
After applying the model to each segment, a simple k-Means clustering is performed on the prototypes to align the segment-level results and compare them against other systems that perform a transcription for the whole meeting, i.e.\ the full \SI{10}{\minute}.
\glsreset{GSS}
\Cref{tab:meeting_results} shows, that using the integrated model from \cref{tab:segment_results}, row \num{8} and using the speaker prototypes for segment alignment ($\boldsymbol{\mu}$) (\cref{tab:meeting_results}, row \num{5}), significantly outperforms a cascaded approach of diarization followed by \gls{GSS} (row \num{1}). Even with subsequent embedding extraction and \gls{SLR} \cite{24_boeddeker_once_more} (row \num{2}), this cascaded approach cannot approach the integrated model.  

\begin{table}[bt]
\setlength{\tabcolsep}{5pt}
    \centering
    \vspace{-1.25em}
    \caption{Comparison of \gls{cpWER} on LibriCSS segments and the full meetings after aligning the segments. \textit{Oracle} denotes the meeting \gls{WER} for an oracle alignment of the speaker permutation across utterances. All configurations use the Nemo ASR model. 
    }
    \label{tab:meeting_results}
    \vspace{-0.5em}
    % Use H to hide a column
    \begin{tabular}{c l H S l S S}
    \toprule
    row & Model & Align & {Segment} & Align & {Meeting} & {Oracle}\\
    \midrule
        1 & VMF $\rightarrow$ GSS \cite{24_boeddeker_once_more} & -- &{--}& --& 12.6 & 9.6 \\
        2 & VMF $\rightarrow$ GSS \cite{24_boeddeker_once_more} & SLR  & {--} & SLR &10.6 & 9.6 \\
    \midrule
        3 & VMF $\rightarrow$ cACGMM & SLR & 14.7 & SLR& 13.9 & 13.3\\ 
        4 & VMFcACGMM  & SLR & 5.7 & SLR& 6.2 & 5.4 \\ %/scratch/hpc-prf-
        5 & VMFcACGMM  & $\boldsymbol{\mu}$& 5.7 & $\boldsymbol{\mu}$ & 6.7 & 5.4 \\
        %/scratch/hpc-prf-nt1/cord/models/vmfcacgmm_segments/busy_peach_carp
        6 & VMFcACGMM $\rightarrow$ GSS  &  $\boldsymbol{\mu}$& {--} & $\boldsymbol{\mu}$& 5.5 & 4.7 \\
        \midrule\midrule
         7 & TS-SEP $\rightarrow$ GSS\cite{24_boeddeker_tssep} & -- & {--} &--& 5.6 & 3.5 \\
        8 & TS-SEP $\rightarrow$ GSS\cite{24_boeddeker_tssep} & SLR & {--} & SLR & 3.7 & 3.5 \\
\bottomrule
    \end{tabular}
    \label{tab:my_label}
\end{table}

Compared to the neural network-based TS-SEP pipeline (row \num{7}/\num{8}), the \mbox{\gls{VMFcACGMM}} loses by more than \SI{1}{\percent} absolute even without SLR and an additional \gls{GSS} postprocessing (row \num{6}) only serves to reduce the difference. 
However, TS-VAD and its extension TS-SEP heavily rely on large amounts of in-domain training data, as was experienced in a CHiME-7 submission \cite{23_boeddeker_chime7}. There, only limited in-domain training data was available, and TS-VAD performed poorly compared to its performance on LibriCSS. 
On the contrary, the \gls{VMFcACGMM} is a strong meeting processing system that can be applied without access to neither in-domain nor paired, i.e.\ corrupted and clean, training data. 
% The processed segments can also directly be aligned without an additional postprocessing component like SLR \cite{24_boeddeker_once_more}. 
For comparison, a TS-SEP system designed for LibriCSS can not be directly used on the DipCo meeting corpus \cite{19_segbroeck_dipco}, a subset of the CHiME-7/8 challenge. To our knowledge, all CHiME challenge participants required training on the DipCo data to design a working system, whereas the \gls{VMFcACGMM} from \cref{tab:meeting_results}  achieves a \gls{cpWER} of 
\SI{52.1}{\percent} without modification or fine-tuning. This is on par with the CHiME-8 Nemo baseline system \cite{24_cornell_chime8} of \SI{54.6}{\percent} which heavily relies on using in-domain training data.

\section{Conclusions}
\label{sec:conclusion}
In this work, we proposed the integration of two statistical mixture models for the simultaneous enhancement and diarization of meeting data. It allows for a robust speech enhancement on meeting segments while at the same time inferring the number of active speakers. 
By using the speaker prototypes of the spectral model to solve the segment permutation problem, 
the model allows a segment-by-segment processing of arbitrarily long meetings. 
Unlike TS-VAD and other neural network-based approaches, the proposed system requires neither in-domain nor paired training data. This is important for in-the-wild meeting transcription, where domain-specific paired data is unavailable for training.
In the future, we will focus on improving speaker embedding quality, thus reducing the complexity of speaker counting and segment alignment.

\section{Acknowledgements}
Computational Resources were provided by BMBF/NHR/PC2.
Christoph  Boeddeker was funded by DFG, project no.\ 448568305.
\pagebreak
\balance
%-------------------------------------------
\bibliographystyle{IEEEbib}
\bibliography{refs}

\end{document}